\documentclass[aps,pra,twocolumn,letterpaper,superscriptaddress,nofootinbib,showpacs]{revtex4-1}
\usepackage{times,mathtools,amssymb,array,sidecap}
\usepackage[hyperfigures,colorlinks=true,allcolors=blue]{hyperref}
\usepackage[utf8]{inputenc}
\usepackage{graphicx}
\usepackage{xfrac}
\usepackage{amsmath,bm}
\usepackage{color}

\newcommand{\z}{\hat{\mathbf{z}}}
\newcommand{\T}{\mathbf{T}}
\newcommand{\N}{\mathbf{N}}
\newcommand{\B}{\mathbf{B}}
\newcommand{\Y}{\mathbf{Y}}
\newcommand{\F}{\mathbf{F}}
\newcommand{\sinc}{\mathrm{sinc}}
\newcommand*{\citen}[1]{%
  \begingroup
    \romannumeral-`\x 
    \setcitestyle{numbers}%
    \cite{#1}%
  \endgroup   
}

\begin{document}

\title{Contractile actuation and dynamical gel assembly of paramagnetic filaments in fast precessing fields}

\author{Joshua M Dempster}
\affiliation{Northwestern University Department of Physics and Astronomy, 2145 Sheridan Road F165, Evanston, Illinois 60208}
\author{Pablo V\'{a}zquez-Montejo}
\affiliation{Northwestern University Department of Materials Science and Engineering, 2220 Campus Drive, Cook Hall 20136, Evanston, Illinois 60208}
\author{Monica Olvera de la Cruz}
\affiliation{Northwestern University Department of Materials Science and Engineering, 2220 Campus Drive, Cook Hall 20136, Evanston, Illinois 60208}
\affiliation{Northwestern University Department of Physics and Astronomy, 2145 Sheridan Road F165, Evanston, Illinois 60208}

\begin{abstract}
Flexible superparamagnetic filaments are studied under the influence of fast precessing magnetic fields using simulations and a continuum approximation analysis. We find that individual filaments can be made to exert controllable tensile forces along the precession axis. These forces are exploited for microscopic actuation. In bulk, the filaments can be rapidly assembled into different configurations whose material properties depend on the field parameters. The precession frequency affects filament aggregation and conformation by changing the net torques on the filament ends. Using a time-dependent precession angle allows considerable freedom in choosing properties for filament aggregates. As an example, we design a field that twists chains together to dynamically assemble a self-healing gel.
\end{abstract}

\maketitle
\section{Introduction}
Despite a history stretching over five decades, magnetic colloids continue to find new applications. Paramagnetic beads are highly attractive bases for dynamic materials due to the relative ease and precision with which researchers can control magnetic fields in many media. Magnetic colloids are being used for medical tasks such as drug delivery \cite{drugdelivery}, tissue scaffolding \cite{tissue2}, image contrast \cite{contrastagents}, and tumor reduction \cite{hyperthermia}. More exotic functions include self-assembling swimmers \cite{swimmers1,swimmers3}, walkers \cite{walkingribbon1,walkingribbon2}, grabbers \cite{asters}, and self-healing membranes \cite{magicangle}.
\\
Both medical \cite{bettermed} and non-medical \cite{lumpactuator,wireflow1,wireflow2} applications  benefit from using particles with extended aspect ratios. 1D chains of magnetic colloids, or magnetic filaments, are of particular interest on account of their rich behavior arising from the interplay of their elastic and magnetic properties \cite{wei2016}. Chain synthesis techniques have advanced steadily over almost two decades \cite{firstchains}, and a wide variety of bead-linking methods are now available to scientists \cite{squishchains, chainbundle, onepot, blockingtemp, ferromagnetic, hollowchain, branching, fattyassembly, anisotropicassembly, streptavidinlink}. In recent years a variety of research teams have studied chain behaviors such as actuation via bending \cite{bending, magnetictorque, coreshell}, buckling transitions \cite{overview1}, desynchronization in slowly rotating fields \cite{rotation}, swimming \cite{swimmers2}, and pumping \cite{pump, pumptheory}.
\\
Most work to date has focused on static \cite{Lusebrink2016, Cerda2016} or relatively slow field changes with single filaments \cite{overview2}. In this work, we combine analytics and simulations to study the behavior of isolated and bulk filaments in dynamic magnetic fields. Our goal is not to characterize the equilibrium ensemble of bulk filaments, as thermodynamic results will depend strongly on the microscopic details of the magnetic filaments. Instead, we focus on exploiting dynamic effects common to many types of filaments in order to produce desired behaviors. In the case of single filaments under fast precession, we find that for large precession angles the filaments naturally form helices with harmonic potentials. These helices can be used for well-controlled contraction at the microscale. For multi-chain ensembles, we show that aggregation behavior is a function of two effects. The dominant effect is the effective interaction of beads in the bulk, which is determined by the precession angle. The second effect is a frequency-dependent residual torque on chain ends. Combining these effects with more complex field choices, such as precession with a dynamically changing precession angle, enables a much greater variety of behaviors. We illustrate these possibilities by driving chains to twist together into an effectively cross-linked gel.

\section{Model}
For computational purposes, we treat the chains as collections of hard, spherical magnetic dipoles bound to their neighbors by flexible but inextensible bonds. The flexibility of the chain is governed by the bending rigidity $\epsilon$ through the harmonic angle potential
\begin{equation}
	U_B (\delta\theta) = \frac{\epsilon}{2} (\delta\theta)^2 \,,
	\label{eq:theta}
\end{equation}
where $\delta\theta$ is the angle formed by a bead and its two neighbors. In addition to hard-sphere interactions, beads also interact via the dipole-dipole potential
\begin{equation}
	U_{ij} = \frac{1}{r_{ij}^3} \left( \bm{\mu}_i \cdot \bm{\mu}_j - \frac{3}{r_{ij}^2} (\bm{\mu}_i \cdot \mathbf{r}_{ij}) (\bm{\mu}_j \cdot \mathbf{r}_{ij}) \right) \,,
	\label{eq:dipole_general}
\end{equation}
where $\bm{\mu}_i$ is the dipole moment of a bead and $\mathbf{r}_{ij}$ the displacement between beads. In MD simulations we take the beads to be freely rotating spheres with dipole moments of fixed magnitude, appropriate to systems such as Ref. \citen{hollowchain} in a saturating external field. We refer to this below as the computational model of the filaments. In the computational model, all magnetic interactions between dipoles less than 10$\sigma$ apart are considered.

The basic behavior of the filaments is readily described. For the regimes considered here we find that magnetization remains uniform. The dominant component of Eq. (\ref{eq:dipole_general}) is then 
\begin{equation}
	U(\mathbf{r}, \bm{\mu}) = \frac{\mu^2}{r^3} \left( 1 - 3 (\hat{\bm \mu} \cdot \mathbf{\hat{r}})^2 \right) \,.
	\label{eq:dipole}
\end{equation}
For the case of a static field, free filaments will align with the field. Parallel filaments have an $r^{-3}$ repulsive interaction when side by side, and an oscillatory attraction that decays exponentially with a length scale set by bead size. Consequently the filaments will eventually aggregate into ribbons and columns, but on time scales too long to interest us here.
\\
Rapidly changing the field orientation frustrates the filament's alignment with the field. Consider a field precessing around the $z$ axis so that the magnetic moment directions are given by $\hat{\bm \mu} = \cos \omega t \, \sin \beta\, \mathbf{\hat{i}} + \sin \omega t \, \sin \beta\, \mathbf{\hat{j}} + \cos\beta\, \mathbf{\hat{k}}$. If the field precesses with a short period compared to the characteristic time scales for bead translation, we are justified in treating the conformation of the filament as quasistatic with respect to the precession. We may average Eq. (\ref{eq:dipole}) over the precession period to obtain the quasistatic magnetic energy
\begin{equation}
	U(r, \beta, \alpha) = \frac{\mu^2}{r^3} \left (1 - 3 \left( \cos^2{\beta}\cos^2{\alpha} + \frac{1}{2} \sin^2{\beta} \sin^2{\alpha}\right) \right)\,,
	\label{eq:precess}
\end{equation}
where $\beta$ is the angle of the magnetic field with its precession axis $z$, and $\alpha$ is the angle of the displacement vector with $z$. For small values of $\beta$, Eq. (\ref{eq:precess}) drives configurations similar to a static field aligned with $z$. For $\beta$ values near $\pi/2$, dipoles prefer to form sheets in the $x-y$ plane, and repel in the $z$ direction \cite{Martin2000}. At the magic angle $\beta_m = \cos^{-1}\sqrt{\sfrac{1}{3}}$, the $r^{-3}$ potential vanishes and interactions are dominated by dipole correlations that scale as $r^{-6}$. Reference \citen{magicangle} discusses the behavior of free colloids at the magic angle in depth. In the present work, we verified that filaments fail to aggregate or align at the magic angle for the parameters used in our system. Instead, we observed dominant hard-sphere interactions. We conclude that the $r^{-6}$ potential is weak compared to thermal forces and does not appreciably contribute to conformations or aggregate formation in the regimes studied here
\\
We assume that the bending modulus $\epsilon$ is large compared to the thermal energy $kT$ and magnetic energy $\mu^2\sigma^{-3}$, so that the chain follows a smooth curve with small $\delta\theta$ everywhere. With this assumption we can now develop a continuum model of the filaments, which is complementary to the computational model above. Let the unit vector $\mathbf{T}(s)$ be the tangent of the filament curve at bead $s$, or equivalently the displacement vector pointing from $s$ to the next bead $s+ds$. Denoting bead diameter with $\sigma$, we write the interaction between neighboring beads as 
\begin{align}
	U_{NN} = \frac{M(\beta)}{2}T_z^2\,, \quad
	M(\beta) = -\frac{9\mu^2}{\sigma^3} \left( \cos^2 \beta -\frac{1}{3} \right) \,,
	\label{eq:NN}
\end{align}
where $T_z = \T \cdot \z = \cos \alpha$. The next-nearest neighbor interactions have leading prefactor $\sfrac{1}{8}$. An expansion in $\delta \theta$ shows that the leading term is simply a rescaling of the nearest-neighbor energy by $1+\sfrac{1}{8}$. The next term is proportional to $\sfrac{3}{16} \mu^2 \sigma^{-3} (\delta \theta)^2$. For the regimes considered here, this term is negligible compared to $U_B$. Accordingly we neglect interactions beyond nearest-neighbor for the continuum model.
\\
We may now write the zero-$T$ Hamiltonian for a single static strand of length $L$ forming a continuous curve $\mathbf{x}(s)$:  
\begin{equation} \label{eq:continuum}
	H = \int_0^L ds\, \left( \frac{\epsilon}{2} (\mathbf{T}')^2 + \frac{M(\beta)}{2} (T_z)^2 \right) \,,
\end{equation}
where the first term gives the bending energy and the second gives the magnetic energy.In the following, we rescale lengths in terms of the bead diameter $\sigma$, v.g. $s \rightarrow s/\sigma$ and the total length $L$ corresponds to the number of beads in the filament. Equation (\ref{eq:continuum}) is similar to previous continuum studies of magnetic filaments, but generalized to the case of fast precession \cite{overview1}. Despite their  simplifications, continuum models well describe experimental results for magnetic filaments in static fields, including U and S-shaped hairpin turns, \cite{overview2}. In the following sections we will use MD simulations of the computational model with a Langevin thermostat to both complement and verify the continuum model in the case of rapid precession and to bound its applicability.

\section{Results}

\begin{figure}
	\centering
	\includegraphics[width=.95\columnwidth]{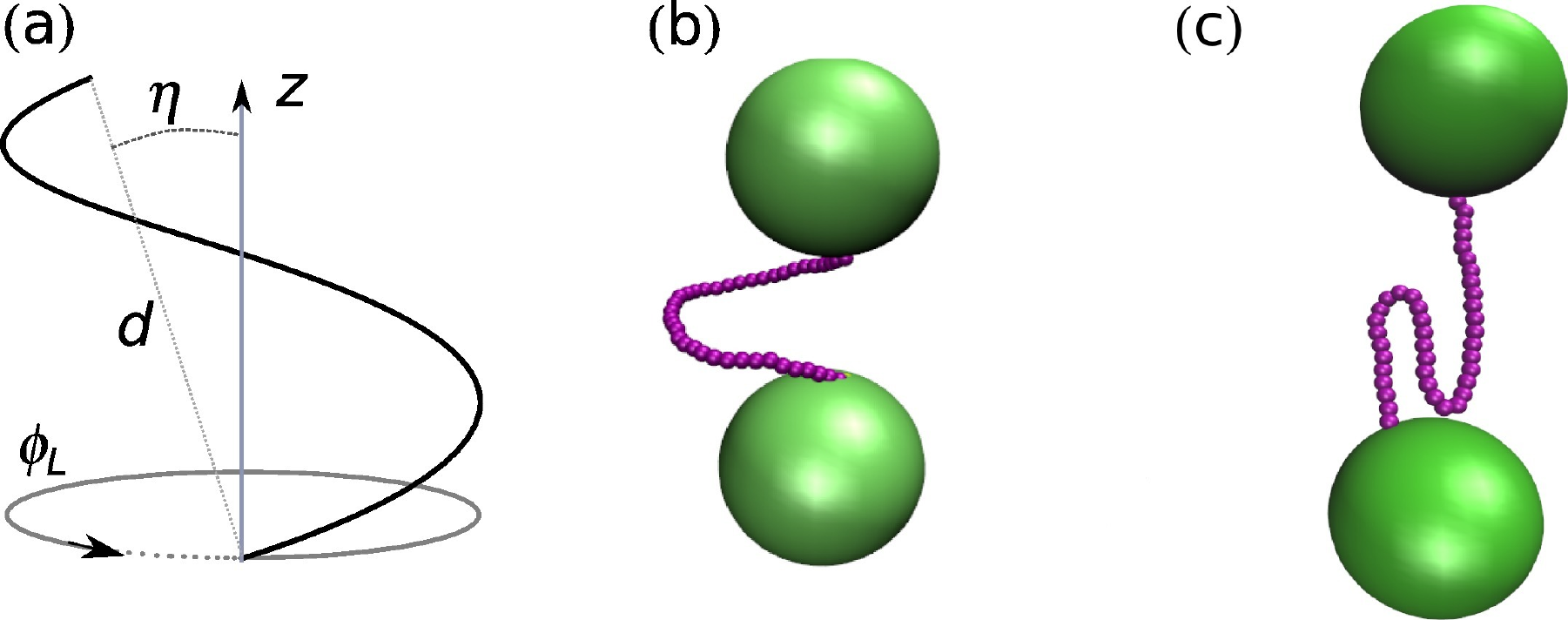}
	\caption{Helical contraction. (a) Schematic of the boundary conditions on a contracting helix. (b) Frame from MD simulation using LAMMPS with a 40 bead magnetic filament connecting two inert test masses. Under the influence of a precessing field with $M > 0$, the helix attempts to contract to the $x-y$ plane. Here $\epsilon = 40$, $\mu^2\sigma^{-3} = 4 kT$, and $\beta=\pi / 2$. (c) If $M < 0$, the filament attempts to align with the $z$ axis and forms planar solutions. Both bending and magnetic energy costs are localized to hairpin turns, so the state is metastable.}
\end{figure}  
The combination of the magnetic and bending energies can be exploited for controllable actuation. Suppose that each end of the filament is anchored to two test loads a distance $d$ apart. In Appendix \ref{sec:Helices} we show that there are two families of solutions for this Hamiltonian with fixed boundaries: helices about the $z$ axis [Figs. 1a and 1b], and planar curves that include the $z$ axis [Fig. 1c]. We find from simulations that the one-coil helices are the stable solutions for positive $M$ ($\beta > \beta_m$). For negative $M$, ($\beta < \beta_m$) the preferred solutions are planar with hairpins that concentrate bending energy, similar to the hairpins that occur for free ends in Ref. \citen{overview1}. 
\\
The total energy of the helix is given for arbitrary initial conditions in Appendix \ref{sec:Helices}. Let us take the ideal case, where the start- and end-points of the chain are aligned with the $z$ axis. For a helix, $T_z = d/L$ and $(\mathbf{T}')^2 = 4\pi^2L^{-2}(1-d^2/L^2)$. Integrating Eq. (\ref{eq:continuum}) we get 
\begin{equation}\label{eq:spring}
	H = \frac{k}{2} d^2 + \frac{2\pi^2\epsilon}{L} \,, \quad 	k = \frac{M(\beta)}{L} - \frac{4\pi^2\epsilon}{L^3}\,.
\end{equation}	
This potential is harmonic in the separation of the end points $d$ for any (physical) value of $d$.  Thus, the helix behaves as a spring with a force constant $k$ that can be adjusted by simply by changing the precession angle $\beta$. If $k$ is positive, the harmonic favors contraction. This offers remarkably precise control over actuation for a given angle.
\\
We verified the formation of harmonic contractile helices with MD simulations in the Large-scale Atomic/Molecular Massively Parallel Simulator (LAMMPS) software package. Details regarding simulation methods can be found in Appendix \ref{sec:mdsim}. Movie 1 in Supplemental Material illustrates the results.

\begin{figure*}
	\centering
	\includegraphics[width=.99\textwidth]{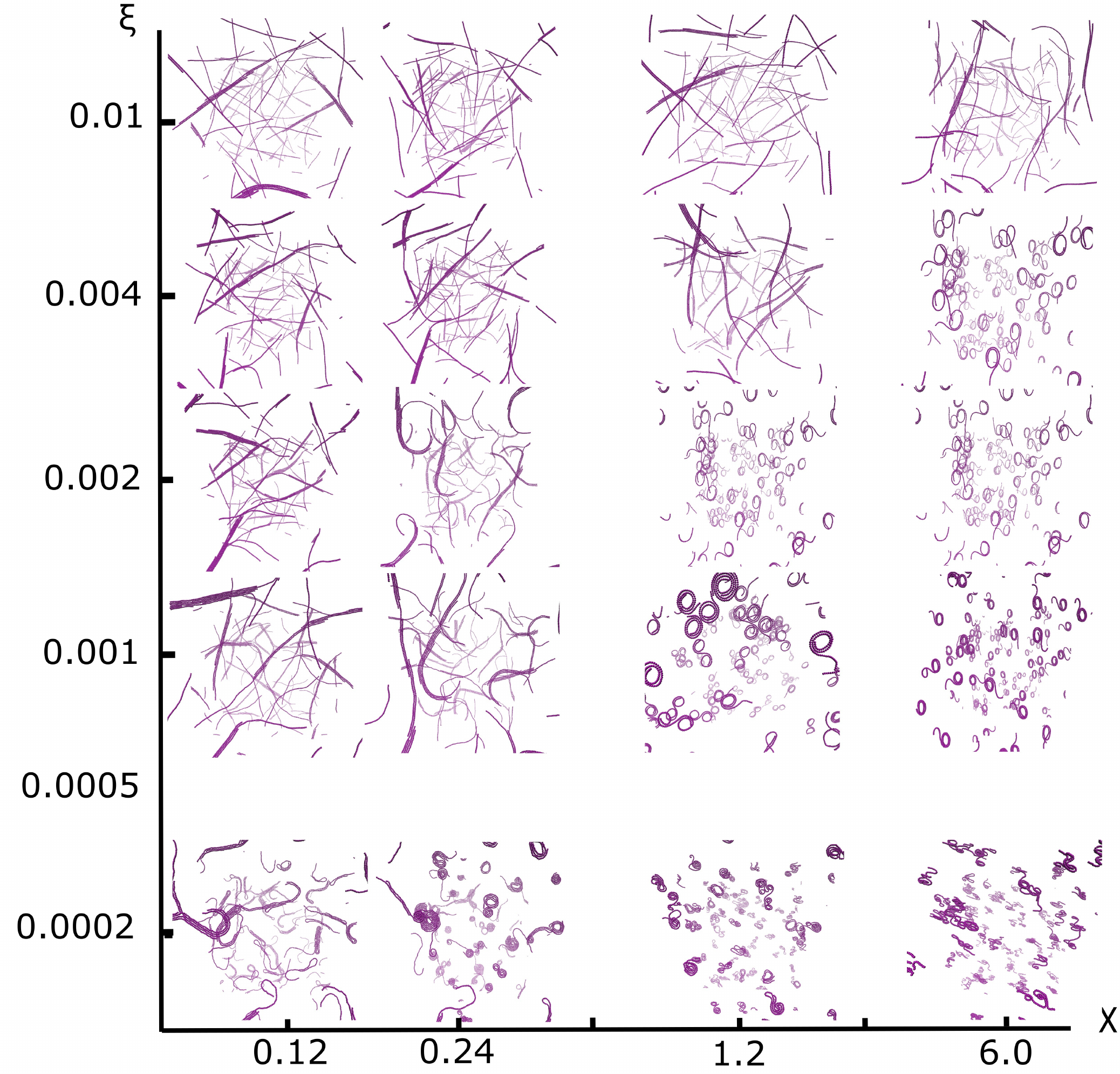}
	\caption{Effects of precession frequency and bending modulus on aggregation for 80-bead filaments with $\mu^2 \sigma^{-3} = 4kT$ and $\gamma \sigma=10$ (Lennard-Jones units). For sufficiently low frequencies and bending moduli, the torque on filament ends forces filaments to aggregate in metastable spirals instead of branching networks.}
\end{figure*}
\vskip0pc \noindent
We turn to the case that the filament ends are not fixed, so that it is possible for multiple filaments to aggregate. If $\beta < \beta_m$, the filaments align with the precession axis in a manner similar to free dipolar colloids. If $\beta > \beta_m$, the filaments arrange themselves in the plane orthogonal to precession and readily aggregate. In this second case, their conformation and aggregation behavior depend on the speed of precession $\omega$. Here we consider the high-(but finite-) frequency regime and determine the minimum necessary frequency for the validity of the Eq. (\ref{eq:precess}). 
\\
For high frequencies the static term given by Eq. (\ref{eq:precess}) remains the dominant energy for individual colloids, but additional force terms arise when the chain is permitted to move in response to finite-frequency precession. For simplicity, consider the case that $\beta=\sfrac{\pi}{2}$, {\it i.e.} that the magnetic field rotates entirely in the plane of the filament. The path of the filament is parametrized by the angle $\psi(s)$ its tangent forms with some axis in the $x-y$ plane. The angle between the tangent and the field is instantaneously $\omega t -\psi$, and we find the nearest-neighbor force $\mathbf{F}_{NN} = -\nabla U(\mathbf{r}, \mathbf{\mu})$ in cylindrical coordinates:
\begin{equation}
\mathbf{F}_{NN} = 3\frac{\mu^2}{\sigma^4} \left[ \sin (2\omega t -2\psi)\mathbf{N} + (1-3 \cos^2 (\omega t - \psi))\T \right] \,,
	\label{eq:FNN}
\end{equation}
where $\mathbf{N}$ is the unit vector in the plane perpendicular to $\T$. For beads in the middle of the filament, the net force is proportional to $\delta \psi$ and amounts to an oscillating adjustment to the bending rigidity $\epsilon$. However, the two end beads have a finite force. Discarding the force in the $\T$ direction, and including the viscosity of the medium, the net force on the ends of a straight filament is given by
\begin{equation}
	\mathbf{F}_{end} = 3\frac{\mu^2}{\sigma^4} \sin (2(\omega - \dot{\psi})t)\mathbf{N} -\gamma \dot{\psi} \mathbf{N} \,,
	\label{eq:Fend}
\end{equation}
where $\gamma$ is the viscous drag coefficient. The magnetic force oscillates back and forth as the field rotates. We are interested in the cycle-averaged motion
\begin{equation}
	\langle \dot{\psi} \rangle = \frac{\omega}{\pi} \int_0^\frac{\pi}{\omega} dt \dot{\psi} \,,
\end{equation}
In a Brownian regime, the instantaneous total force $\mathbf{F_{end}}$ is zero. Further, in the high-frequency regime the motion of the filament in a single precession cycle is small. We write $\psi=\psi_0+\delta \psi$, and set $\psi_0=0$ without loss of generality. Expanding the force to linear order in $\delta \psi$, we find
\begin{equation}
	\dot{\psi} = \frac{1}{\tau} \big ( \sin (2\omega t) - 2\delta \psi \cos (2\omega t) \big)
	\label{eq:psidot}
\end{equation}
where we have introduced the magnetoviscous timescale $\tau = 1/3  \gamma \,\mu^{-2} \sigma^{4}$. On averaging over one period of the motion, the left term vanishes. The right term may be integrated by parts:
\begin{equation}
	\langle \dot{\psi} \rangle = \frac{ 1}{ \pi \tau} \int_0^\frac{\pi}{\omega} dt  \dot{\psi} \sin (2\omega t)
	\label{eq:recursion}
\end{equation}
One may now recursively substitute Eq. (\ref{eq:psidot}) into Eq. (\ref{eq:recursion}), integrate the first term, and integrate the second term by parts. Concretely, the first such iteration generates
\begin{align}
	\langle \dot{\psi} \rangle &= \frac{ 1}{ \pi \tau} \int_0^\frac{\pi}{\omega} dt \frac{1}{\tau} \big ( \sin (2\omega t) - 2\delta \psi \cos (2\omega t) \big) \sin (2\omega t) \nonumber \\
	&= \frac{ 1}{ \pi \tau} \left( \frac{\pi}{2\omega\tau} + \frac{1}{4\omega \tau} \int_0^\frac{\pi}{\omega} dt  \dot{\psi} \sin (4\omega t) \right) \nonumber \\
	&= \frac{1}{\tau} \left( \frac{1}{2\omega \tau} + \mathcal{O} (\omega \tau) ^ {-2} \right)
	\label{eq:first_recurse}
\end{align}
The net motion of the filament ends against the drag force of the solvent may be identified with an average magnetic force $\langle \mathbf{F}_{NN} \rangle = \gamma \dot{\psi} \N$, which vanishes with $(\omega \tau)^{-1}$. This net force generates a magnetic torque that tends to wind the two ends of the filament in curves of opposite chirality. The magnetic torque competes with the bending modulus to determine the conformation of the filament aggregates. For large values of $\omega$, the bending modulus dominates and filaments aggregate in branching networks. For smaller values, the magnetic end force succeeds in forcing the filament ends to curve until they make contact with the body of the filament. Magnetic attractions between beads then cause the filament to roll into a spiral, and separate spirals aggregate in compact clusters.  Figure 2 illustrates the resulting aggregate metastable ``phases'', characterized in terms of the inverse magnetoelastic parameter $\xi = \epsilon / (ML^2)$ and magnetically normalized end force $\chi = M / (\gamma \omega \sigma)$, with $M= 3 \mu^2/\sigma^3$, as generated by the MD model. 
\\
For large values of the bending modulus the end force never rolls the filament in a spiral, instead rotating the entire filament coherently (top right). For small bending moduli, the filaments tend to naturally roll and fold even at high frequencies (bottom left). At low frequencies, the quasistatic approximation for the filament bulk fails and the filaments form rapidly rotating compact shapes (bottom right). For intermediate values of $\xi$, there is a continuous transition from branching to spiral conformations in the region $500 \le \chi/\xi \le 800$.  In all cases, filaments were initially thermalized with no magnetic interactions. Note that since the aggregates are metastable, alternative initial conditions will alter the diagram.
\\
It is now possible to bound the validity of Eq. (\ref{eq:precess}) at $\chi \le 1$. For 100 nm-diameter filaments in room temperature water with the parameters quoted in Fig. 2, this requires a precession frequency of 1 kHz. The required frequency falls with increasing diameter as $\sigma^{-3}$. 
\begin{figure*}
	\centering
	\includegraphics[width=.99\textwidth]{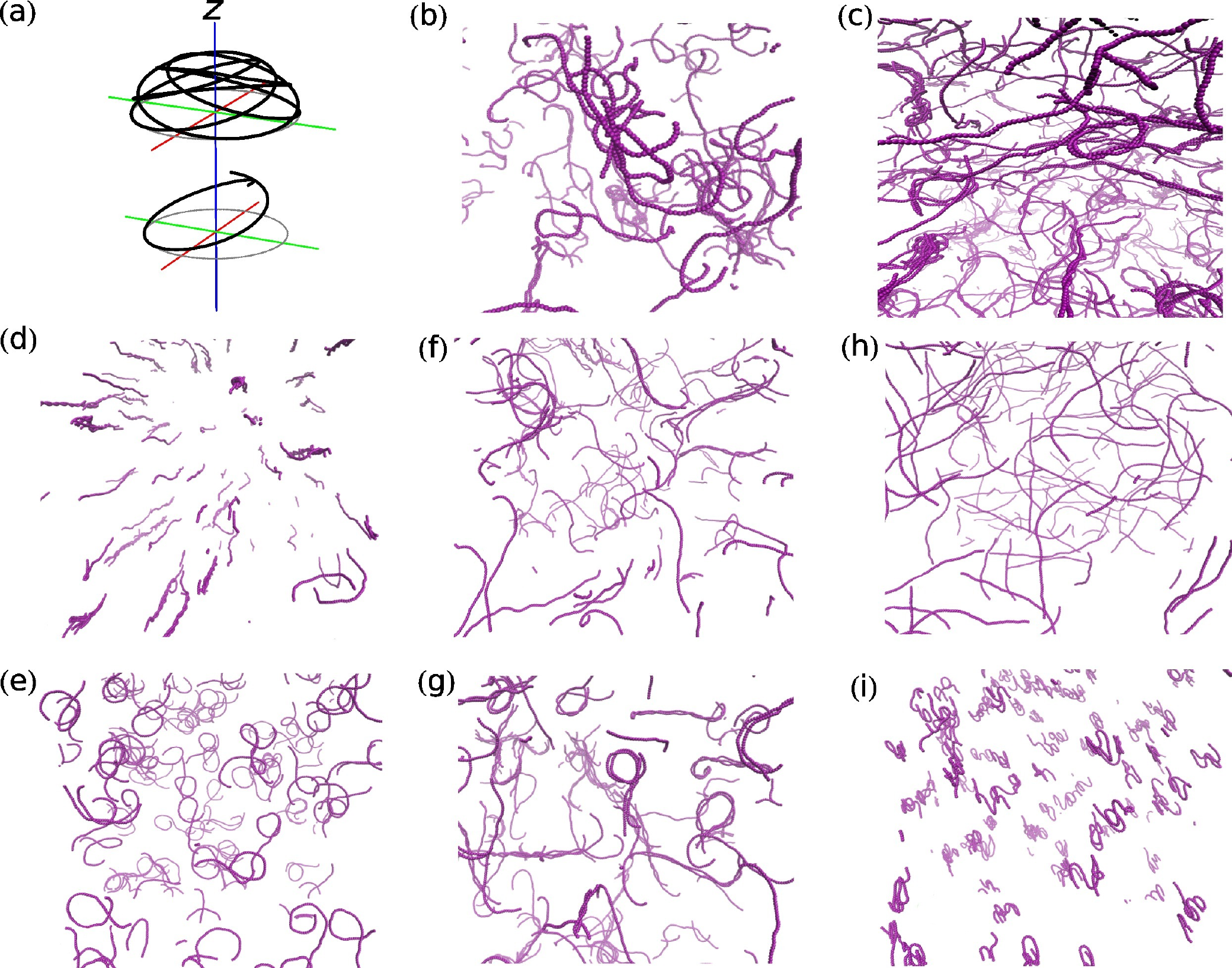}
	\caption{Achieving twisting with a magnetic drive. (a) Path traced by the driving field with $\omega=2.5$, $\omega_2=1.0$, and $\beta_0=0.5$ over $t \in [0, 4\pi]i$ (top) and $[0, \pi]$ (bottom). (b) The twisted cross-links created by this drive with the parameters $\mu=2.0$ and $\epsilon = 80.0$. Further simulation details are given in the SM. (c) Increasing chain length and system density creates a system-spanning network of cross-linked filaments. The succeeding panels are the results of altering the magnetic drive parameters as noted: (d) $\beta_0 = 0.4$; (e) $\beta_0 = 0.6$; (f) $\omega_2 = 2.0$; (g) $\omega_2 = 0.5$; (h) $\omega = 25$, $\omega_2 = 10$; (i) $\omega = 0.25$, $\omega_2 = 0.1$.}
	\label{Fig:3}
\end{figure*}
More complex aggregates are possible with a more general form of driving field than simple precession. A truly arbitrary field opens an enormous design space. Here, we confine ourselves to illustrating how one may design the field to achieve a desired end state. We select as our target material a self-assembling 3D gel in which strands have twisted around one another to effectively cross-link. Such a gel is much stronger than one that relies on magnetic attractions alone, and is of special interest due to its dynamic nature and potential for self-healing.
\\
For an isotropic gel, we require $\langle M \rangle \approx 0$ while retaining a nonzero $r^{-3}$ attraction between filaments. To achieve this we use a precessing field as before, but with a time-dependent precession angle:
\begin{equation}
	\beta(t) = \frac{\pi}{2} - \left( \frac{\pi}{2} - \beta_0 \right) \cos^2 \omega_2 t
	\label{eq:beta}
\end{equation}
and set $\beta_0$ so that the precession angle roughly averages to the magic angle $\beta_m$, or $\beta_0 \approx 0.5$. Obtaining the desired twisting behavior is more challenging. The end points of the filaments should trace a circular pattern around the filament axis, and should do so regardless of how the filament is aligned in the laboratory frame. Therefore, the field must trace open loops on the unit sphere at frequencies low enough for the filament ends to respond. Guided by the preceding frequency analysis of the filament response to finite frequencies, we choose $\xi=0.001$ and $\chi \approx 1$. In order to ensure an isotropic gel, each filament end must respond to the precession regardless of filament orientation. We therefore choose $\omega = 2.5$, $\omega_2 = 1$, so that the field forms a series of open loops with different orientations. Figure 3(a) illustrates the path the field traces on the unit sphere with these parameters.
\\
We test these parameters in MD simulations of 80-bead filaments with $\epsilon=40$, $\mu=2$, and volume fraction is $0.001$. Figure 3(b) illustrates the successful ``spun'' cross-links, but the system density is too low to form a system-spanning network. We achieve such a network in Fig. 3(c) by increasing chain length to 160 and volume fraction to 0.0018. 
\\
We also verify the effects of changing the field parameters in the low-density system. Setting $\beta_0$ too high or too low creates a net bias in the chain bulk interactions towards planar or linear conformations, preventing gelation [Figs. 3(d)-3(e)]. If $\omega_2/\omega$ is too large or too small, the chain ends do not exhibit strong circular motion and chains do not spin together into a gel [Fig. 3(f)-3(g)]. Large values for $\omega$ and $\omega_2$ lead to weak torques on chain ends and no chain aggregation [Fig. 3(h)], while small values violate our assumption that we can time-average magnetic interactions in the bulk. In this case, chains form compact dynamic shapes as the entire chain responds to changes in the field [Fig. 3i].

\section{Conclusion}
In this paper, we presented analytic and computational studies of superparamagnetic filaments under the influence of a rapidly precessing field. Single filaments with their ends attached to loads can be used as microscopic actuators, with the strength of contraction determined by the angle of precession. Collections of free chains are capable of complex configurations. We found that these configurations are determined primarily by two effects: the desired orientation of beads in the chain bulk, which is fixed by the precession angle, and the residual torque on chain ends which vanishes as the inverse square of the precession frequency. We can exploit the combination of these effects to generate particular desired materials and behavior. Our work demonstrates the power of magnetic filaments as the basis for dynamic designer materials, in which desired properties can be created and destroyed in real time. We note that in the case of driven systems, metastable structures may depend on the initial configuration, a feature worth of exploring. Extending the present discussion with exact equilibrium configurations for filaments in multiple precession regimes is a subject for future research. 

\section*{Acknowledgements} 

This work was supported by the Center for Bio-Inspired Energy Science (CBES), which is an Energy Frontier Research Center funded by the U.S. Department of Energy, Office of Science, Office of Basic Energy Sciences under Award No. DE-SC0000989.

\begin{appendix}
 
\section{Helices in a precessing field} \label{sec:Helices}

Here we show that helices minimize the energy of the curve given by the elastic and magnetic contributions. Let $\Y(s)$ be the position vector of the curve parametrized by arc-length $s$ and of total length $L$, with given start and end points for the chain, $\Y(0) = \Y_0$ and $\Y(L)=\Y_L$, subject to the physical constraint that they are not more than a distance $L$ apart, i.e., $d = |\Y_L - \Y_0| < L$.
The Frenet-Serret (FS) frame adapted to the curve is composed by $\T(s)= \Y'(s)$ the tangent vector, $\N(s)$ the principal normal
and $\B(s) = \T(s) \times \N(s)$ the binormal. The FS formulas
describe how the frame rotates along the curve:
\begin{equation}
	\begin{pmatrix} \T ' \\ \N ' \\ \B ' \end{pmatrix} = 
	\begin{pmatrix} 
		0 & \kappa & 0 \\
		-\kappa & 0 & \tau \\
		0 & -\tau & 0 
	\end{pmatrix}
	\begin{pmatrix} \T \\ \N \\ \B \end{pmatrix}\,,
\end{equation}
where $\kappa = \T' \cdot \N$ is the curvature and $\tau = \N' \cdot \B$ is the torsion. If $\kappa=0$, the curve is a straight line; if $\tau=0$, the curve lies in a plane.
\\
In order to minimize the total energy $H$ defined in Eq. (\ref{eq:continuum}), it is convenient to consider the effective energy, \cite{Auxiliary} :
\begin{equation} \label{eq:efcontinuumham}
	H_E = H + \int_0^L ds\, \left( \F \cdot (\T-\mathbf{\Y}') + \frac{\lambda}{2}(\T^2-1) \right) \,,
\end{equation}
where $\F$ is an auxiliary vector field acting as a Lagrange multiplier implementing the definition of the tangent vector as the arc-length derivative of the position vector \cite{Sphereconf, Envbias}, whereas $\lambda$ is another Lagrange multiplier implementing rigid bonds, thus constraining the distance between beads.
\\
The Euler-Lagrange equation for $\Y$ yields the conservation law $\mathbf{F}'=0$. 
Using the FS equations and decomposing $\z$ in the FS basis, the Euler-Lagrange equation for $\mathbf{T}$ yields $\F$ spanned in the FS basis
\begin{eqnarray}
	\mathbf{F} &=& -\left(\epsilon \kappa^2 + M T_z^2 +\lambda \right)  \T + \left(\epsilon \kappa' - M T_z N_z \right) \N \nonumber \\
	&&+	\left(\epsilon \kappa \tau - M T_z B_z \right) \B \,.
\end{eqnarray}
Differentiating $\F$, using again the FS formulas and projecting onto the tangent, we get 
\begin{equation}
\mathbf{F}' \cdot \T =  -	\left(\epsilon \kappa^2 + M T_z^2 + \lambda \right)' - \kappa (\epsilon \kappa'- M T_zN_z) 
	\label{eq:FpT}
\end{equation}
Recognizing that $\kappa N_z = T_z'$, we can write the second term as a total derivative--a consequence of the reparametrization invariance of the energy \cite{Hamforcurves}---allowing us to integrate and solve for $\lambda$:
\begin{equation}
	\lambda = -\frac{1}{2}(3 \epsilon \kappa^2 + M T_z^2) + c \,,
	\label{eq:lambda}
\end{equation}
where $c$ is a constant of integration, which can be regarded as a line tension controlling the total length.
We can now write
\begin{eqnarray}
\mathbf{F} &=& \left(\frac{\epsilon}{2} \kappa^2 + \frac{M}{2} T_z^2 - c \right)  \T + \left(\epsilon \kappa' - M T_z N_z \right) \N \nonumber \\
&&+	\left(\epsilon \kappa \tau - M T_z B_z \right) \B\,.
\end{eqnarray}
$-\F$ is the external force on the curve. The EL equations correspond to the projections onto the two normals 
\begin{subequations} \label{eq:FpNB}
\begin{eqnarray}
\mathbf{F}' \cdot \N &=& \epsilon \left( \kappa'' + \kappa \left(\frac{\kappa^2}{2}-\tau^2\right) \right) \nonumber \\
&& + \kappa\left(M \left(\frac{T_z^2}{2}- N_z^2 \right) - c \right) =0 \,, \label{eq:FpN}\\
\F' \cdot \B &=&	\epsilon\left( \kappa \tau' + 2 \tau  \kappa' \right) - M \kappa N_z B_z =0\,. \label{eq:FpB}
\end{eqnarray}
\end{subequations}
Equation (\ref{eq:FpB}) can be satisfied when $\tau$ and $B_z$ vanish, which corresponds to configurations lying on a single plane containing the $z$ axis. A detailed analysis of these planar curves, determined by solving Eq. (\ref{eq:FpN}) under appropriate boundary conditions, will be discussed elsewhere. 
\\
Since $\F$ is conserved, its magnitude is constant
\begin{eqnarray}
	F^2 &=& \left( \frac{\epsilon}{2} \kappa^2 + \frac{M}{2} T_z^2 - c\right)^2 + (\epsilon \kappa' - MT_zN_z)^2 \nonumber \\
	&&+(\epsilon \kappa \tau-MT_zB_z)^2\,.
	\label{eq:F2}
\end{eqnarray}
This provides a first integral of Eqs. (\ref{eq:FpNB}). The trivial way to satisfy this equation is to set all the quantities on the right hand side to be constant. Moreover, for constant $\kappa$, $\tau$, $T_z$ and $B_z$, Eq. (\ref{eq:FpB}) is satisfied if $N_z=0$, which corresponds to a helix winding around the $z$ axis. In this case Eq. (\ref{eq:FpN}) determines the constant $c$.
\\
In cylindrical coordinates $(\rho,\phi,z)$, the position vector of a $Z$-aligned helix centered on the origin with radius $\rho$ and pitch $p = 2 \pi \xi$\footnote{$\xi$ can be expressed as $\xi = \rho \tan \gamma$, with $\gamma$ the angle that the tangent makes with the azimuthal direction $\hat{\bm \phi}$, so it measures the slanting of the helix.} can be spanned with respect to the cylindrical basis $\{\hat{\bm \rho}, \hat{\bm \phi}, \hat{\bf z}\}$ as $\Y(s) =  \rho \hat{\bm \rho} + \xi \phi \hat{\bf z}$. The tangent vector is $\T = (\rho \hat{\bm \phi} + \xi \hat{\bf z})/\sqrt{\rho^2 + \xi^2}$. The FS curvature and torsion of the helix are $\kappa = \rho /(\rho^2 + \xi^2)$ and $\tau = \xi /(\rho^2 + \xi^2)$.
\\
To complete the characterization of the helical segment, besides $\rho$ and $\xi$, the total azimuthal angle $\Phi$ between the two ends has to be specified, so $\phi \in (0, \Phi]$. In this manner also the height difference is determined, $h= \xi \Phi$ and $z \in (0, h]$. The total length $L$ of the helix segment and distance $d$ between the two ends are given in terms of these three parameters by
\begin{equation} \label{eq:defLd}
L = \Phi\sqrt{\rho^2 + \xi^2}\,, \quad d = \sqrt{4 \rho^2 \sin \frac{\Phi}{2} + \xi^2 \Phi^2} 
\end{equation}
Alternatively, the helical segment can be described in terms of $L$, $d$ and the angle $\eta$ that the displacement vector between the two ends makes with the helical axis. To this end we start by expressing $\rho$ and $\xi$ through the relations
\begin{equation}  \label{eq:xietarel}
 2 \rho \sin \frac{\Phi}{2} = d \sin \eta \,, \quad h = \xi \Phi = d \cos \eta  \,.
\end{equation}
Combining these expression with those for $L$ and $d$ in Eq. (\ref{eq:defLd}), we find that the total angle $\Phi$ is given in terms of $L$, $d$ and $\eta$ by the equation
\begin{equation}
	\sinc{\frac{\Phi}{2}} = \frac{\sin \eta}{\sqrt{\left(\frac{L}{d}\right)^2 - \cos^2{\eta}}}\, ,
	\label{eq:phiL}
\end{equation}
where $\sinc \, x = \sin x/x$. Since this is a transcendental equation it has to be solved numerically. Once $\Phi$ is known, $\rho$ and $\xi$ can be determined from Eqs. (\ref{eq:xietarel}).
\\
The bending and magnetic energies of the helical segment are given by 
\begin{equation} \label{eq:EBEMhlx}
H_B =  \frac{\epsilon}{2} \frac{\Phi^2}{L}\left(1 - \left(\frac{h}{L}\right)^2\right)  \,, \quad 
H_M = \frac{M(\beta)}{2}\, \frac{h^2}{L}\,. 
\end{equation}
Clearly, the bending energy is minimized by reducing $\Phi$ and increasing $h$, so that the helix with one or fewer windings and higher pitch is preferred. By contrast the magnetic energy is minimized by reducing $h$, or equivalently the pitch of the helix, tending to form approximately circular loops on the plane orthogonal to the direction of precession. We see that the competition of the bending and magnetic energies can be exploited for controllable, repeatable actuation: the first tends to straighten the helix, whereas the latter tends to collapse it.
\\
In the special case that the helix completes one period $\Phi = 2\pi$, ($\eta=0$ and $h=d$) the total energy of the chain reduces to Eq. (\ref{eq:spring}).

\section{MD Simulation} \label{sec:mdsim}

We used the open source code LAMMPS to perform molecular dynamics simulations of beads. System units were chosen such that bead diameters, bead masses, the magnetic constant $\mu_0/(4\pi)$, and thermal energies were unitary. Beads in a field were modeled as point dipoles of fixed magnitude $2.0$, with hard sphere interactions given by truncated Lennard-Jones potentials, linear bonds with force constant $k=500$, and a Langevin thermostat with damping set to 0.1 so that the drag constant on each bead is ten times its mass. This ensures an overdamped behavior. Stiffness was set by a harmonic angular potential as described in the text. The external field was of magnitude $2000$, so its interaction strength with the dipoles was $4000 kT$. 
\\
For the helices in Fig. 1(b) and the ESI movie, chains of length 40 beads were anchored to fixed points on the surface of load spheres with diameter 10 and mass 1000. We attempted both fixed and unfixed tangent boundary conditions. The fixed tangent boundary condition is enforced by applying the angular potential to the angle between the chain tangent at the surface of the load sphere and the radial vector from the center of the load sphere to the chain attachment point. Switching between boundary conditions did not observably alter the resulting dynamics. Figure 2(b) and the ESI movie use the fixed tangent condition.
\\
The conformations of Fig. 2 were generated with 100 chains of length 80 beads. The volume density for the system was 0.002 in system units.
\\
The conformations of Fig. 3 use the same system parameters as Fig. 2, except that the two cases the demonstrate twisting (Figs. \ref{Fig:3}(b) and \ref{Fig:3}(c)) were repeated with 400 chains of 120 beads to ensure the twist-linked gel remains intact at larger system sizes. In this case the density was 0.0035. 
\\
Various conditions, including the spirals of Fig. 2(a) and the twisted gel of Fig. 3(b), were tested with lattice-Boltzmann hydrodynamics. Lattice spacing was set to bead size, and viscosity and density were set to 1 in system units. We did not observe any significant changes, but cannot guarantee that sufficiently strong hydrodynamic effects will not disrupt the results reported here. 
 
\end{appendix}

\bibliography{magnetic_polymers} 
\bibliographystyle{apsrev4-1} 

\end{document}